\documentclass{aa}
\usepackage{natbib}
\usepackage{graphicx}
\usepackage{txfonts}

\begin{document}

\title{Planets in binary systems: is the present configuration indicative of 
the formation process?}
\subtitle{}

\titlerunning{Planets in binary systems}

\author{F. Marzari
        \inst{1}
        \and
        M. Barbieri
        \inst{2}
        }

   \offprints{F. Marzari}

   \institute{
              Dipartimento di Fisica, University of Padova, Via Marzolo 8,
              35131 Padova, Italy\\
              \email{marzari@pd.infn.it}
         \and
             LAM, Traverse du Siphon, BP 8, Les Trois Luc,
             13376 Marseille Cedex 12, France\\
             \email{mauro.barbieri@oamp.fr}
             }

   \date{Received 16 January 2007 ; accepted 26 January 2007}

\abstract 
{}
{The present dynamical configuration of planets in 
binary star systems may not reflect their formation process
since the binary orbit may have changed in the past 
after the planet formation 
process was completed. An observed binary system may have been
part of a former hierarchical triple that became unstable after the planets
completed their growth around the primary star. 
Alternatively, in a dense stellar 
environment even a single stellar encounter between the 
star pair and 
a singleton may singificantly alter 
the binary orbit.  In both cases the planets we
observe at present would have formed when the dynamical 
environment was different from the presently observed one.}
{We have numerically integrated the trajectories of 
the stars (binary plus singleton) and of test planets 
to investigate the abovementioned mechanisms. The orbits of 
the planets have been computed to test the 
survival of a planetary system around the primary 
during the chaotic phase of the stars.}
{Our
simulations show that
the circumstellar environment
during planetary formation around the primary 
was gravitationally less perturbed when the binary was part
of a hierarchical triple because the binary was
necessarely wider and, possibly, less eccentric. This circumstance has
consequences for the planetary system in terms
of orbital spacing, eccentricity, and mass of the individual
planets. Even in the case of a single stellar encounter 
the present appearance of a planetary system in a binary 
may significantly differ from what it had
while planet formation was ongoing. However, while in
the case of instability of a triple the trend is always 
towards a tighter and more eccentric binary system, when a single stellar
encounter affects the system the orbit of the binary
can become wider and be circularized.}
{
Modelling the formation of a planetary system around a binary
is a potentially complex task and an effort has to be made
to look into its present dynamics
for traces of a possible chaotic past.}

\keywords{Planets and satellites: formation; Methods: N--body simulations;
         Celestial mechanics}

\maketitle

\section{Introduction}

Among the many extrasolar planets discovered so far, more than 40
are gas giants detected in binary or even triple
star systems. This is not an unexpected result since it is
known that about half of solar type stars near the sun 
have a companion and there is evidence that the 
binary frequency might be higher among young stars 
\citep{rezi,kole,nghe}.
Most of the binary systems hosting a planet have a 
wide separation
ranging from 100 to 1000 AU (see \citealt{desibar} for a compilation of 
binary systems with planets). However, there are some systems 
like $\gamma$ Cephei, Gl 86 and HD 41004A where the separation 
between the two stars is of the order of 20 AU or less. Planetary
formation in these systems is perturbed to a significnat 
extent by the gravitational
influence of the companion star possibly altering the process. 
\cite{the04} have shown that the formation of the planet in 
$\gamma$ Cephei via core--accretion \citep{pollo} is unclear.
The migration in the massive gaseous disk  required
to form a planet within 2-2.5 AU from the star leads to 
an orbit much closer to the star than that observed. On the other 
hand, \cite{boss} has shown that in hydrodynamical models with 
artificial viscosity the presence of a companion star may 
help to trigger the formation of giant planets by disk instability.  

The different models for planetary formation are applied 
considering 
that the present orbital configuration of the 
system star--planet(s)--star was unchanged. However, there are
cases where the orbit of the companion star may have been 
modified between the time when planetary formation was ongoing 
and the present. If the binary separation was wider or the eccentricity
lower, the protoplanetary disk would have been a more quiet environment 
promoting the formation of planets.
There are two possible 
mechanisms that may have altered a past binary dynamical 
configuration into the presently observed one. 

\begin{itemize}

\item We assume that the observed  binary system 
was in the past part of an unstable 
larger multiple system (triple or more). We will concentrate 
here on hierarchical triples, the most
common multiple system, in which the inner binary is
orbited by a third star on a wider orbit.  The dynamical
evolution
of these systems may be complex in particular in the
presence of high inclinations between the binary and the 
singleton. We may have systems whose parameters 
change nearly periodically by  large amounts that
are stable over long timescales.
However, if the triple is dynamically unstable and the 
system becomes chaotic, the most frequent outcome 
(stellar collisions are rarer) is the 
ejection of a star leaving a binary system with smaller 
separation, because of orbital energy conservation,
and higher eccentricity. The planet in 
a circumstellar orbit around the 
main star of the pair has a time window to form in a
quiet disk before the onset of dynamical instability
of the triple system and subsequent modification of the 
original binary orbit into a more eccentric and closer one. 

\item If the binary system formed in a densely populated cluster
it is likely that the star couple underwent an encounter with another 
star. During the close stellar passage the orbital parameters of the 
binary system are permanently modified leading to a different 
configuration star--planet(s)--star.  Even in this case the 
mean trend is towards more eccentric orbits and, on
average, to an orbital shrinking of the binary pair \citep{heg,lar0}.
\end{itemize}

Both these mechanims, besides altering the binary configuration,
have also the potential to destabilize
the circumprimary
orbits of potential planets. However, as we will see in the following, 
inner circumprimary orbits, like that of the planet
in $\gamma$ Cephei, are little or not affected by the stellar 
encounters in most cases.  

In Sect. 2 of this paper we will describe how the decay of a
hierarchical triple star system affects the orbital parameters
of the binary system that was part of it and of a putative planetary
system around the primary star of the system.
To simplify the approach we will model only coplanar 
hierarchical systems even if inclined triples may be more
common because of frequent encounters in star forming regions
early on in stellar evolution. 
This restriction to coplanarity is suggested by the minor complexity
of the star dynamics, compared to the high inclined systems, that 
allows us to ouline clearly the effects of the stellar instability
on the planet environment. We 
will also investigate 
to what extent the close encounters between the secondary stars 
perturb a planetary system around the main star. In Sect. 3 
we will analyse the effects of stellar encounters on a binary 
system with planets. We will present  
selected cases that illustrate the 
difficulty of interpreting the formation of a planet in a binary
system in light of its present orbital paramters. We will not perform
a full exploration of the parameter space for the two mechanisms
described in Sect. 2 and 3 which would be beyond of the scope of
this paper.  

\section{Jumping stars and planet survival}

According to \citep{rep,kro,lar0} most binaries and even single stars 
originate from the decay of multiple star systems. The wide
range of orbital parameters observed among binaries might be 
a consequence of the chaotic dynamics of the primordial multiple
systems. Dynamical interactions among the members of the system
cause an exchange of energy and angular momentum often resulting in
the ejection of a star from the system. This may occur in the early 
stages of the system evolution or it may take some time for instability
to build up. We will consider a scenario in which a triple 
star system is stable long enough to allow planet formation around
the primary star of the binary pair. The subsequent destabilization
of the triple, because of the onset of chaotic behaviour, will 
end with the ejection of one star. The binary couple will have 
orbital parameters that are at present different from those of the primordial 
binary in which planetary formation occurred.  
In some cases, during the stellar encounters an exchange can also occur 
between the original companion star and the outer one leading to 
a new mass ratio of the binary system. 

\subsection{The numerical algorithm}

To analyse the dynamical evolution of triple star systems 
and of planetary orbits around the primary star we have 
assembled a numerical model of the hierarchical three--body
system. We define two initial osculating Keplerian orbits 
for the triple, an inner one 
for the binary pair $m_1$ and $m_2$ and an outer one, defined 
in the center of mass of the pair, for the single star $m_3$.
Ten additional circular orbits for massless bodies 
starting at 1 AU from the primary
star and extending by equal steps up to 10 AU are also computed. 
Their evolution allows us to evaluate to what extent the 
chaotic phase of stellar encounters affects a putative planetary
system around the primary star. 
The evolution of the system is calculated with the numerical 
integrator RADAU \citep{rad} which properly handles close encounters
between the massive bodies. 

The parameter space of the hierarchical three--body problem
is rather wide since in principle there are 12 degrees of freedom:
the six parameters of the binary pair and the six ones of 
the singleton. In this work we restrict ourself to a limited number of 
significant test cases showing how a detached binary system has its 
orbital elements changed when the triple is disintegrated 
and the single star escapes to infinity. A similar behaviour occurs 
also in non--coplanar systems when the dynamics becomes chaotic, 
however we will perform a full exploration of the 
inclined case in a forthcoming paper.
We start with the
stars in prograde and coplanar orbits with the semimajor
axis of the binary $a_b$ fixed to 35 AU. The semimajor axis 
of the single star $a_s$ is instead regularly sampled starting within
the empirical stability limit derived by \cite{egki}. We consider 
different values of $e_b$, the eccentricity of the binary, and
of $e_s$, the eccentricity of the singleton. Because of the chaotic
character of the dynamics, for each set of $(a_b, a_s, e_b, e_s)$ 
we perform 30 different simulations with random choices of the 
orbital angles. In this demonstrative set of simulations 
the masses of the stars are
1, 0.4 and 0.4 solar masses, respectively.

\subsection{The onset of instability}

\begin{figure}
\resizebox{\hsize}{!}{\includegraphics{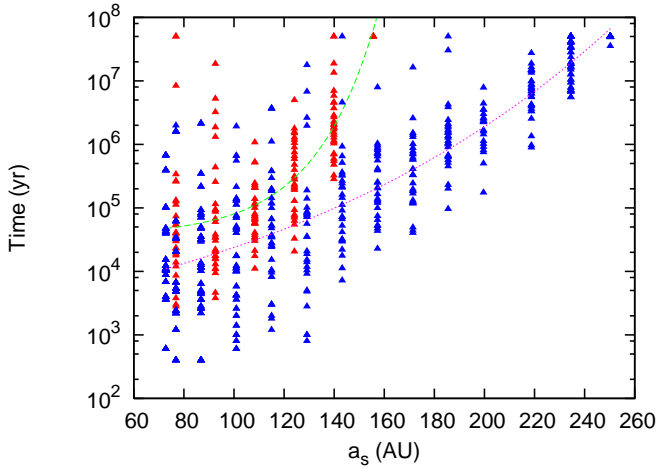}}
\caption[]{Time of the first close encounter between the binary and 
the singleton 
vs. $a_s$, the initial semimajor axis of the singleton. The red dots
mark the cases with binary eccentricity $e_b = 0.2$ and 
singleton eccentricity $e_s = 0.2$, the blue dots 
those with $e_b = 0.4$ and $e_s = 0.4$. The continuous lines are least squares
exponential 
fits to the data.}
\label{f1}
\end{figure}

In Fig.\ref{f1} we plot the timescale $t_{in}$ required for the onset of 
instability, and then close encounters between the single star and the binary,
as a function 
of the semimajor axis of the singleton. In the first case 
(red triangles) the initial orbital 
eccentricities $e_b$ and $e_s$ are both set to 0.2. There is
a trend towards higher values of $t_{in}$ for larger values of 
$a_s$ that can be roughly described 
by an exponential law. Since we stop the simulations after 50 Myr because
of too long integration times, the curve is biased at the 
end of the fit by vales that are all equal to 50 Myr.
Despite this limitation of the fit, at a first sight the analytical
dynamical stability threshold
\citep{egki,egki2}, located at about  $a_s = 150 AU$ for the 
singleton, appears to be in good agreement with our numerical
results if one looks at the 
fast growing trend of the instability time when $a_s$ is
larger than 150 AU. In the second case shown in Fig.\ref{f1} (blue
triangles) we have incresed both the eccentricities $e_b$ and $e_s$ to 0.4. 
The instability onset occurs, as expected, faster than the previous 
case and larger values of $a_s$ are required to have long surviving 
systems. To find systems whose $t_{in}$ is 
longer than 50 Myr we have to go beyond $a_s \sim 250$ AU. 
The stability limit set by the formula of \citep{egki,egki2}
is at about 235 AU and it seems to be slightly
less accurate at high eccentricities of the stars.

According to \cite{boss} disk instability can form gas giant
planets in a few hundred years. In this scenario the time required 
by the gravitational perturbations among the stars to build up 
dynamical instability is not a crucial parameter. When $a_s$ is beyond 
70 AU the chaotic behaviour onset occurs after a few thousand 
years giving enough time for planets to form. Eventually, 
if the planetary system is chaotic after its formation 
\citep{wema1,wema2,wema3,rafo} it might have enough time to evolve into
a stable state before the star system becomes in turn chaotic.  
In the core accretion theory \citep{pollo}, the formation of gas giant planets
requires a few million years. By inspecting Fig.\ref{f1} we expect that only
systems in the upper part of the plot are dynamically stable long
enough to grant a quiet environment for planets to grow around 
the primary star. However, even in some 
cases where the singleton is close to the binary, the dynamics turns 
chaotic only after some million years. 

\begin{figure}
\resizebox{\hsize}{!}{\includegraphics{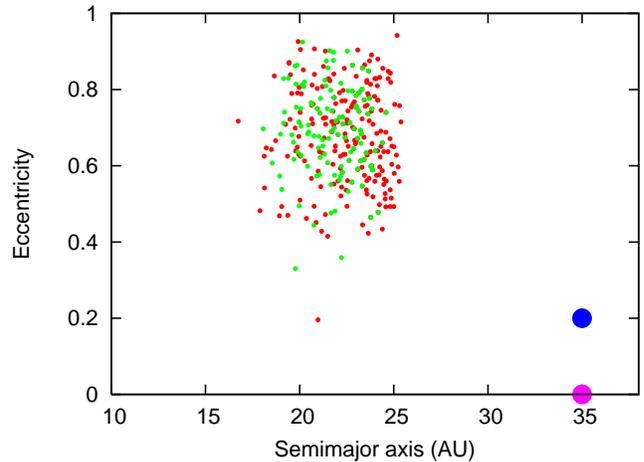}}
\caption[]{Distribution of the binary orbital elements after the third 
star of the system has escaped on a hyperbolic orbit. The red dots 
label binaries originated from triples with $e_b = 0.2$ and $e_s = 0.2$, 
the green dots those from triples with $e_b = 0$ and $e_s = 0$. The 
large dots in the plot represent the initial binary system.}
\label{f2}
\end{figure}

In Fig.\ref{f2} we show the final orbital distribution in the 
$(a_b,e_b)$ plane of the left--over binaries. While the initial 
values of $e_b$ and $e_s$ have a strong influence on 
the instability time $t_{in}$, they appear non influential 
for the final 
orbital distribution of the binaries. The two different 
distributions shown in Fig.\ref{f2} correspond to 
$e_b = 0.2$ and $e_s = 0.2$ (red dots) while 
the green dots mark systems where both the binary and the singleton are
on circular orbits. There is no significant difference between the 
two distributions that overlap in the $(a_b,e_b)$ plane.
The final semimajor axis 
$a_b$ is substantially smaller than the initial value 
($a_{b0}=35$ AU) as
expected by orbital energy conservation as one of the stars escapes on 
a positive energy hyerbolic orbit. The spreading 
of $a_b$
depends both on the different initial values of $a_s$ and 
on the amount of energy taken away by the singleton.
The final values of $e_b$ are randomly distributed and are concentrated at high
eccentricities, in most cases larger than 0.4. The results of additional 
simulations with sampled values of $e_s$ and $e_b$ show a similar 
distributions in the final orbital distribution of the binary
star system.  

\begin{figure}
\resizebox{\hsize}{!}{\includegraphics[angle=-90]{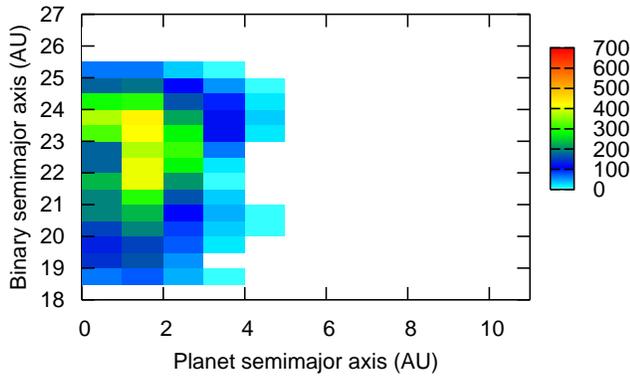}}
\caption[]{Histogram illustrating the number of systems 
in our simulations that, at the end of the chaotic phase,
fall into bins 
in $a_b$, the binary separation, and $a_p$, the semimajor axis of 
the outer planetary orbit that survives the stellar chaotic phase.
}
\label{f3}
\end{figure}

In Fig.\ref{f3} we map the values of the semimajor axis $a_p$ of the 
outer planetary orbit that remains
stable after the chaotic phase 
vs. the final value of $a_b$ in the case with $e_b = 0.2$ and $e_s = 0.2$. 
The color codes are linked to the number of cases that fall in each 
bin in $a_p$ (x--axis) and $a_b$ (y--axis).
The repeated encounters between the stars set a limit of about 
4 AU from the primary star within which planetary orbits can survive. 
Beyond that value of $a_p$, the gravitational disturbances produced during the chaotic 
evolution of the two outer stars destabilize any planetary body. 
It is remarkable that a significant number of binary systems with final 
values of $a_b$ around 19 AU and high eccentricity, preserve the 
planetary system, or part of it, that formed 
when the binary was part of a triple system with $a_{b0}=35$ AU. 
$\gamma$ Cephei, Gl 86 and HD 41004A might well belong to this
class of systems. In 1\% of our simulations even the innermost planetary
orbit we considered ($a_p = 1$ AU) is perturbed during the stellar 
chaotic phase and the planet is eventually ejected from the 
system. This implies that a fraction of planetary systems forming 
around binaries, part of a triple in the early stages of their evolution,
may be totally destabilized during the stellar encounters.

Our study is not exaustive in terms of exploration of the parameter space,
however it gives a clear idea of the dynamical evolution of 
the system.  It also suggests caution when modelling planetary formation
in an observed binary system by simply adopting its present orbital configuration.
That configuration might indeed be the outcome of a complex 
and unpredictable chaotic evolution.  

\section{Modification of star--planet--star configurations after 
a stellar encounter}

Within stellar clusters, close stellar encounters
may disrupt binaries or, more frequently,  abruptly change 
their orbital elements \citep{kro}. This mechanism is suspected 
to be one of the possible causes of the low frequency of binaries
among low--mass field stars compared to that of young low--mass 
stars in star--forming associations \citep{duch}. If a planet or
a full planetary system formed
around one of the stars of a binary, when a stellar
encounter occurs the dynamical configuration of the system
may be significantly changed \citep{stel}. We may see  
either of the
following events:
\begin{itemize}
\item The binary system is destroyed and from then on
the surviving planets orbit 
a single star. In this case, the dynamical configuration we 
observe at present is not indicative of the formation 
process because of the large changes caused by the stellar 
encounter. 
\item The binary system survives the encounter but 
its orbital parameters and those of the planets 
are strongly altered. 
\end{itemize}

In both cases, any attempt to model the formation of an 
observed planetary system would face the problem of 
discriminating which dynamical or physical features of the system 
are due to the formation process and which are related  
to the stellar encounter.

\begin{figure}
\center
\includegraphics{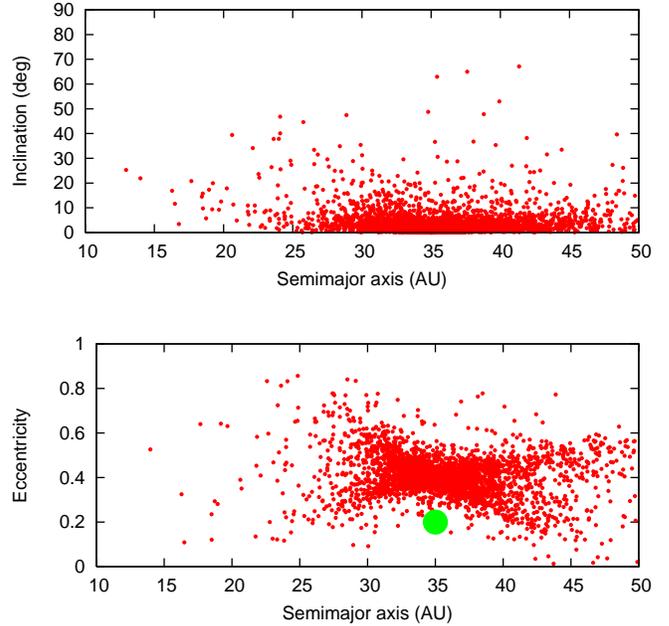}
\caption[]{Distribution of the eccentricity and inclination 
of the binary system
after the hyperbolic stellar encounter. 
All the systems shown in the plot allowed the 
survival of at least one planet around the primary star of the system.
The large dot represents the initial binary configuration.}
\label{f4}
\end{figure}

In this Section we investigate to what extent the dynamical environment of 
a planet orbiting a star in a binary system is affected by a stellar 
encounter that does not disrupt the binary. We adopt the 
same numerical model described in the previous section, but
the singleton is now set on a hyperbolic orbit approaching the 
binary on a plane tilted by $45^{\circ}$ with respect to the 
binary orbital plane. We randomly sample the impact parameter 
and eccentricity of 
the hyperbolic trajectory and we look at the orbital configuration of
the final system star--planet--star when the encounter is over. 
In Fig.\ref{f4} we show the 
orbital distribution in the (a,e) and (a,i) planes
of the binary systems after the stellar 
encounter. A significant number of 
systems 
has a post--encounter semimajor axis significantly smaller than the initial 
one ($a_{b0}=35 AU$) while during the decay of a triple system the final
$a_b$ cannot be lower than a value fixed by the conservation of
the gravitational energy of the initially bounded system.  

\begin{figure}
\resizebox{\hsize}{!}{\includegraphics[angle=-90]{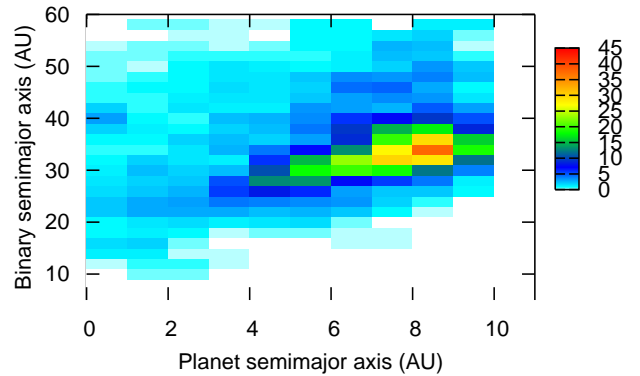}}
\caption[]{
Plot showing the semimajor axis $a_p$ of the outer planetary orbit that
survives the stellar encounter vs. 
the post--encounter semimajor axis of the companion star
$a_b$.
To be compared with Fig.\ref{f3}.}
\label{f5}
\end{figure}

The impact of a stellar encounter on planetary formation models 
is important: the planet(s) may have formed when the companion star
was farther away and then less effective in perturbing the 
protoplanetary disk. For the core--accretion model \citep{pollo}, 
this may be
a crucial condition to allow planetesimal accumulation before 
the onset of strong binary perturbations when the companion 
stars is injected on a closer orbit \citep{the04,the06}.
By inspecting the planetary orbits that 
are still bound to the primary star after the 
encounter shown in Fig.\ref{f5}, 
we observe a distribution very  different from that of 
Fig.\ref{f3}. We still have cases where planetary orbits up to 
3 AU survive the encounter leaving the planets around a closer
binary system than the initial one. When we look at 
$\gamma$ Cephei, Gl 86 and HD 41004A we may be observing
a system that survived a close stellar encounter. 
However, it appears 
that the single encounter with a star on a hyperbolic orbit
has less dramatic consequences on
the stability of a circum--primary planetary system 
than the frequent close 
encounters between bounded stars that lead to the decay of 
a triple. Most of the planetary systems show orbits that 
are still stable up to 10 AU when the stellar encounter is 
over. 
Of course, at later times outer orbits far from the primary
may become unstable because of the long term perturbations of
the companion star \citep{howi}.

While the onset of instability in a triple system always leads
to tighter and more eccentric binary systems, 
by inspecting Fig.\ref{f4} we see systems where the orbit of the
binary couple is broadened and, in a few cases, even circularized 
by the stellar encounter. 
In these cases we know that planetary formation was perturbed 
by the companion 
star in the initial stages, but 
the dynamical environment of the planetary system became significantly 
quieter after the stellar encounter. 
There are systems where the binary companion is moved beyond 
some hundreds of AU from the main star. If the stellar encounter occurred
after the planets formed, we might observe a broad binary system
with a small planetary system around the main star. Trying to understand 
why the orbits of such a planetary system are eccentric or too close to the
star without considering the possibility of a stellar encounter 
would lead to wrong conclusions. 

\section{Conclusions}

Any attempt to model the formation and 
evolution of a planetary system around the component of 
a binary star system should account for the possibility
that the binary orbit has changed 
with time. A dynamical event like the destabilization of
an original triple system or a close stellar encounter can significantly
alter the binary orbit after the formation of planets
around the primary star. A system like 
$\gamma$ Cephei where a planet orbits the main star at about 2 AU 
and is perturbed by a secondary star presently moving 
at about 20 AU might have had a different dynamical configuration when 
the planet formed. Trying to interpret the present configuration of 
the system using the observed orbital and physical parameters may lead to 
misleading deductions on the physical properties of the 
protoplanetary disk that generated the planets. 
Prior to the binary orbit change, the disk may have been more (or less) 
affected by the gravitational pull of the companion star that 
was on a different orbit. The individual masses and 
orbit separations of the planets that formed in this environment
might not be well simulated by introducing in the 
planetary formation model the present parameters of the system. 
Planetesimal accretion, which is the first step of 
terrestrial planet formation and of the core--accretion model 
for giant planet growth may occur differently depending on the
binary orbit and companion star distance. Even gravitational instability
might follow a different path if the companion star was farther
away in the initial phase of the binary life. 

In a few cases a stellar encounter may even cause the
stripping of the companion star from the binary.
The outcome would be a single star with a
planetary system that formed when the star was part of a binary
system and then perturbed by the
gravity field of the companion star.
In addition,   
stellar encounters may even
push the orbit of the companion star out of its original plane 
leading to a significant mutual inclination between the planet
and the companion. This is an efficient mechanism to produce planets
in a Kozai resonance with the secondary star. 

\begin{acknowledgements}
We thank P. Eggleton for his useful comments and suggestions while acting as 
referee of the paper.
\end{acknowledgements}


\begin{thebibliography}{}

\bibitem[Boss(2006)]{boss} Boss, A., 2006, ApJ 641, 1148.
\bibitem[Dalla Stella(2005)]{stel} dalla Stella, A., Marzari, F., 
Barbieri, M., Vanzani, V., Ortolani, S., 2005, 
36th LPSC 2005, n. 1253.
\bibitem[Desidera and Barbieri(2007)]{desibar} Desidera, S. and Barbieri, M.,
2007, A\&A in press.
\bibitem[Duchene(1999)]{duch} Duchene, G., 1999, A\&A 248, 485.
\bibitem[Eggleton and Kiseleva(1995)]{egki} Eggleton, P., and Kiseleva, L.,
1995, ApJ 455, 640--645.
\bibitem[Kiseleva et al.(1996)]{egki2} Kiseleva, L., Aarseth, S., Eggleton, P., and
de al Fuente Marcos, R., 
1996, ASP Conf. S.  90, 433--435.
\bibitem[Ghez et al.(1993)]{nghe} Ghez, A.M., Neugebauer, G., \& Matthews,
K., 1993, AJ 106, 2005.
\bibitem[Heggie (1975)]{heg} Heggie, D.C., 1975,
MNRAS 283, 566.
\bibitem[Holman and Wiegert(1999)]{howi} Holman, M. J., and Wiegert, P. A.,
1999, ApJ 117, 621. 
\bibitem[Kohler and Leinert(1998)]{kole} Kohler, R. \& Leinert, C., 1998,
A\&A 331, 977. 
\bibitem[Kroupa(1995)]{kro} Kroupa, P., 1995, MNRAS 277, 1491.
\bibitem[Larson (2001)]{lar0} Larson, R.B., 2001, in
IAU Symposium n. 200, H. Zinnecker and R.D: Mathieu, Eds.
\bibitem[Marzari and Weidenschilling(2002)]{wema2} Marzari, F. and Weidenschilling, S.J.,
2002, Icarus 156, 570.
\bibitem[Marzari(2005)]{wema3}
Marzari, F., Weidenschilling, S.J., Barbieri, M., Granata, V., 2005,
ApJ 618, 502.
\bibitem[Pollack et al(1996)]{pollo}
Pollack, J. B., Hubickyj, O., Bodenheimer, P., Lissauer, J. J.,
Podolak, M., Greenzweig, Y., 1996, Formation of the Giant Planets by
Concurrent Accretion of Solids and Gas, Icarus 124, 62.
\bibitem[Rasio and Ford(1996)]{rafo} Rasio, F.A. and Ford, E.B., 1996,
Science 274, 954.
\bibitem[Everhart(1985)]{rad} Everhart E., 1985. 
In: Carusi A., Valsecchi G.B. (eds.) Proc. IAU Coll.
83, Dynamics of comets: their origin and evolution. Reidel, 
Dordrecht, p.185
\bibitem[Reipurth and  Zinnecker(1993)]{rezi} Reipurth, B. and Zinnecker, 
H., 1993, A\&A 278, 81.
\bibitem[Reipurth(2000)]{rep} Reipurth, B., 2000, AJ 120, 3177.
\bibitem[Th\'ebault et al.(2004)]{the04} Th\'ebault, P.,
Marzari, F., Scholl, H., Turrini, D., Barbieri, M.,2004,
Planetary formation in the $\gamma$ Cephei system, A\&A, 427, 1097
\bibitem[Th\'ebault et al.(2006)]{the06} Th\'ebault, P.,
Marzari, F., Scholl, H., 2006,
Relative velocities among accreting planetesimals in binary systems:
The circumprimary case, Icarus, 183, 193
\bibitem[Weidenschilling and Marzari(1996)]{wema1} Weidenschilling, S.J. and Marzari, F.,
1996, Nature, 384, 619.



\end{thebibliography}
\end{document}